\documentclass{emulateapj}
\def\Fref#1{Figure~\ref{Fig:#1}}
\def\Tref#1{Table~\ref{Table:#1}}

\newcommand{\altm}  {\altaffilmark}
\newcommand{\kms}   {km~s$^{-1}$}
\newcommand{\mpy}   {mas~yr$^{-1}$} 
\newcommand{\vperp} {V_{\perp}}
\newcommand{\vns}   {V_\mathrm{NS}}
\newcommand{\Mc}    {M_\mathrm{c}}
\newcommand{\Mns}   {M_\mathrm{NS}}
\newcommand{\Msun}  {M_\odot}
\newcommand{\pmra}  {\mu_{\alpha}}
\newcommand{\pmdec} {\mu_{\delta}}

\shorttitle{The Hyperfast Pulsar B1508+55}
\shortauthors{Chatterjee et al.}
\slugcomment{For the Astrophysical Journal Letters}

\begin{document}
\title{Getting Its Kicks: A VLBA Parallax for the Hyperfast Pulsar B1508+55}
\author{
S. Chatterjee\altm{1}, W. H. T. Vlemmings\altm{2,3}, W. F. Brisken\altm{4}, 
T. J. W. Lazio\altm{5}, J. M. Cordes\altm{3}, W. M. Goss\altm{4},
S.~E.~Thorsett\altm{6}, E. B. Fomalont\altm{7}, A. G. Lyne\altm{2},
and M. Kramer\altm{2}  
}
\begin{abstract}
The highest velocity neutron stars establish stringent constraints on
natal kicks, asymmetries in supernova core collapse, and the evolution
of close binary systems.  Here we present the first results of a
long-term pulsar astrometry program using the VLBA.  We measure a
proper motion and parallax for the pulsar B1508+55, leading to
model-independent estimates of its distance
($2.37^{+0.23}_{-0.20}$~kpc) and transverse velocity
($1083^{+103}_{-90}$~\kms), the highest velocity directly measured for
a neutron star.  We trace the pulsar back from its present Galactic
latitude of 52.3\arcdeg\ to a birth site in the Galactic plane near
the Cyg OB associations, and find that it will inevitably escape
the Galaxy.  Binary disruption alone is insufficient to impart the
required birth velocity, and a natal kick is indicated.  A composite
scenario including a large kick along with binary disruption can
plausibly account for the high velocity.
\end{abstract}


\altaffiltext{1}{Jansky Fellow, National Radio Astronomy Observatory;
        and Harvard-Smithsonian Center for Astrophysics, 60 Garden Street,
        Cambridge, MA 02138;  schatterjee@cfa.harvard.edu}
\altaffiltext{2}{Jodrell Bank Observatory, University of Manchester, 
   Macclesfield, Cheshire SK11 9DL, UK.} 
\altaffiltext{3}{Department of Astronomy, Cornell University, Ithaca, NY 14853.}
\altaffiltext{4}{National Radio Astronomy Observatory, P.O. Box O,
   Socorro, NM 87801.} 
\altaffiltext{5}{Naval Research Laboratory, Code 7213, Washington, DC 20375.}
\altaffiltext{6}{Department of Astronomy and Astrophysics, University
   of California, Santa Cruz, CA 95064.}
\altaffiltext{7}{National Radio Astronomy Observatory, 520 Edgemont
        Road, Charlottesville, VA 22903.} 

\section{Introduction}\label{Sec:intro}

The high velocities of young radio pulsars relative to their
progenitor stellar populations have been readily apparent, both in
their proper motions, and through the bow shock nebulae produced in
some cases.  Models to explain these high neutron star (NS) velocities
have invoked the disruption of close binary systems
\citep{b61,ggo70,it96}, natal kicks from asymmetric supernova (SN)
explosions \citep{s70,dc87}, and post-natal acceleration through the
electromagnetic rocket effect \citep{ht75}.  As the pulsar population
velocity distribution has been elucidated \citep[see,
e.g.,][]{acc02,hllk05}, binary disruption and the electromagnetic rocket
effect have required more extreme assumptions to remain viable, and
natal kicks have emerged as the most plausible mechanism for NSs to
gain high velocities.  In turn, observed pulsar velocities have placed
stringent constraints on SN core collapse mechanisms. In simulations,
a natal kick of the correct order of magnitude has been obtained from
hydrodynamic and convective instabilities \citep{bh96,jm96}, and more
exotic mechanisms involving asymmetric neutrino emission in the
presence of strong magnetic fields ($B \gtrsim 10^{15}$~G) have also
been suggested \citep{al99}.

At this point, two caveats are worth careful attention. First, the
current pulsar velocity distribution suffers from severe selection
effects. The stellar progenitors of NSs inhabit the Galactic plane, as
do we, and thus higher velocity pulsars spend less time within our
detection volume than do lower velocity ones. The effect is
exacerbated by pulsar searches that focus more heavily on the
target-rich environment of the Galactic plane.
The extremely successful Parkes Multibeam survey, for example, spans
only $|b|<5\arcdeg$ \citep[e.g.,][]{pmb1}. Thus the observed velocity
distribution is likely to be biased toward lower velocities.

Second, most pulsar distances (and hence velocities) are inferred
from their pulsar dispersion measures (DM $\equiv \int_0^D n_e \, dl$)
and models for the Galactic distribution of electron density ($n_e$)
along the line of sight (\citealt{tc93}, hereafter TC93;
\citealt{cl02}, hereafter NE2001) which have substantial
uncertainties.  For example, the Guitar Nebula pulsar B2224+65
\citep{crl93} is a showcase example of ultra-high transverse
velocities, $\vperp \approx 1640 \,(D/1.9~{\rm kpc})$~\kms, but the
distance uncertainty in the NE2001 model implies that $\vperp \lesssim
1000$~\kms\ cannot be ruled out.  The highest extant model-independent
estimate of a NS velocity is $630 \pm 40$~\kms\ for PSR~B1133+16
\citep{bbgt02}.

Here we present the first hyper-fast pulsar ($V>1000$~\kms) with a
{\em model independent} distance and transverse velocity measurement.
PSR~B1508+55 was part of a recently concluded astrometry program with
the NRAO Very Long Baseline Array (VLBA), in which 27 pulsars were
observed for eight epochs each, spanning 2 years.
This Letter serves to introduce the project, while comprehensive
papers on our astrometric methods and parallax results for $\sim$20
pulsars are in preparation (W. F. Brisken et al., S. Chatterjee et al.).  
Below we report astrometric results for the distance, velocity and
birth site of B1508+55, and discuss the implications of the high
observed velocity.


\section{Observations and Data Analysis}\label{Sec:obs}

\begin{deluxetable}{ll}
\tablecolumns{2}
\tablewidth{0pc} 
\tablecaption{Parameters for PSR B1508+55\label{Table:fit}}
\tablehead{ 
\colhead{Parameter} & \colhead{Value}
}
\startdata 
Epoch & 2003.0 (MJD 52640) \\
$\alpha_0$ & $15^{\mathrm h} 09^{\mathrm m} 25\fs6211$ \\
$\delta_0$ & $55\arcdeg 31\arcmin 32\farcs331$  \\
$\pmra$ (\mpy) & $-73.606 \pm 0.044$ \\
$\pmdec$ (\mpy) & $-62.622 \pm 0.088$ \\
$\pi$ (mas) & $0.415 \pm 0.037$ \\
\\
D (kpc) & $2.37^{+0.23}_{-0.20}$ \\
$\vperp$ (\kms) & $1083^{+103}_{-90}$ \\
$l,b$  &  91.325\arcdeg, 52.287\arcdeg \\
$\mu_l,\mu_b$ (\mpy) & $-6.39, \; 96.06$\\   
\\
Magnetic Field $B$ (G) & $1.95 \times 10^{12}$ \\
Spindown Age (yr) & $2.34 \times 10^6$\\
$D_{\mathrm{DM}}$ (kpc)  & 1.9 (TC93), 0.9 (NE2001) \\
\enddata 
\tablecomments{All astrometric parameters are in the J2000 coordinate
  system.  Absolute positions are referenced to J1510+5702, and have
  an uncertainty $\sim$1~mas.  $\mu_l, \mu_b,$ and $\vperp$ are
  corrected for differential Galactic rotation.}
\end{deluxetable} 

\begin{figure*}[ht]
\plottwo{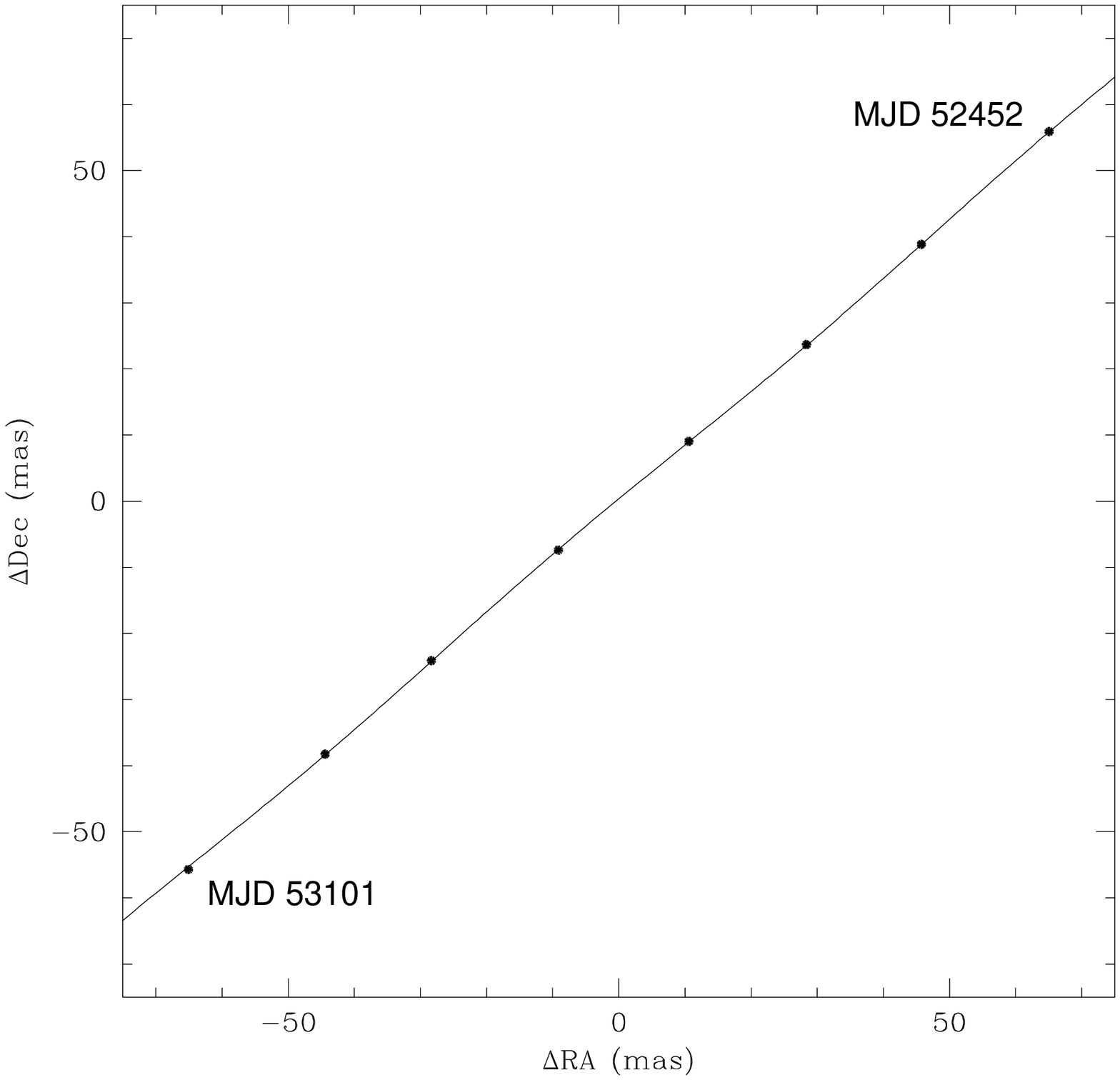}{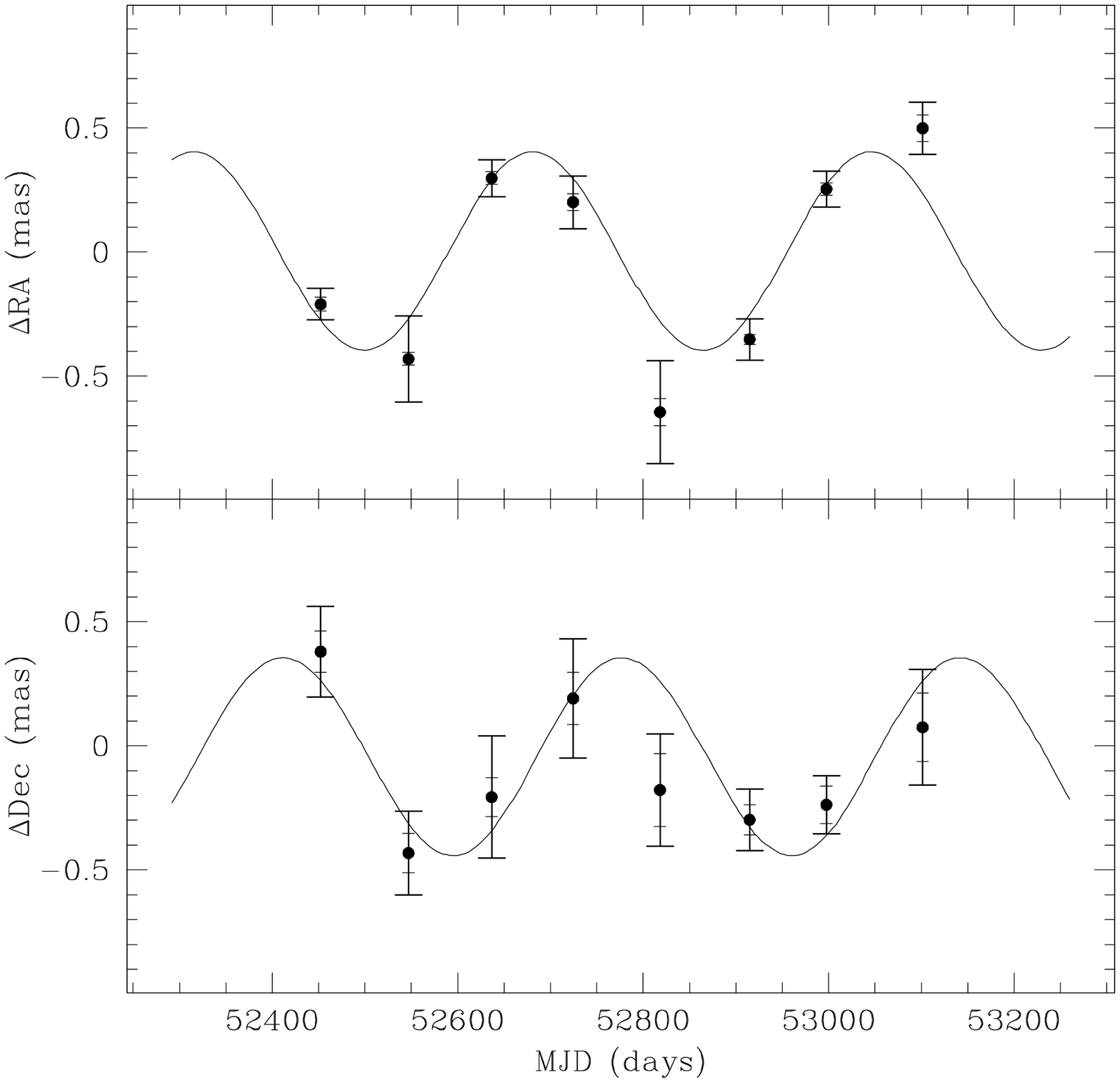}
\caption{Left: The motion of PSR~B1508+55 in Right Ascension and
  Declination, with the best fit proper motion and parallax model
  overplotted. The error estimates for each data point are smaller
  than the size of the points. Right: The parallax signature of
  PSR~B1508+55 in Right Ascension and Declination, after subtracting
  the best-fit proper motion from the astrometric positions. Curves
  corresponding to the best fit parallax $\pi = 0.415$~mas are
  overplotted. The inner error bars indicate the random position
  uncertainties, while the outer error bars indicate the net
  uncertainties (random and systematic, added in quadrature).
}
\label{Fig:pmpi}
\vspace{-6mm}
\end{figure*}

B1508+55 was observed with the VLBA over eight epochs spaced by
approximately 3 months between 2002 June and 2004 April.  As a
trade-off between increasing pulsar flux at lower frequencies and
improved resolution as well as reduced ionospheric effects at higher
frequencies, the observations were conducted between 1.4 and 1.7~GHz,
with four frequency bands of 8~MHz each and dual polarization in each
band.  In order to retain visibility phase-coherence, observations
were phase-referenced by nodding ßbetween the target and a primary
calibrator source (J1510+5702) $\sim 1.5\arcdeg$ away, with a cycle
time of 120~s on target and 90~s on the calibrator.  Residual
calibration errors (primarily from the unmodeled ionosphere over
1.5\arcdeg) are much reduced by employing in-beam calibration
\citep{fgbc99,ccl+01}, which reduces the angular throw and eliminates
the need for time interpolation.  Thus the observations are finally
referenced to a faint ($\sim 7.1$~mJy) extragalactic source
(J151148+554156) only 22.6\arcmin\ away from the pulsar.  The
signal-to-noise ratio (S/N) for the pulsar was boosted by gating the
correlator on at the expected times of arrival of pulses, using
current pulse timing solutions obtained for each epoch from ongoing
observations at the Jodrell Bank Observatory.

Data reduction was conducted in {AIPS} using a customized pipeline
(W. F. Brisken et al.\ 2005, in preparation).  Briefly, it included
amplitude calibration based on system temperatures at each antenna,
followed by calibration of the visibility phases, rates, and delays
based on the observed visibility phases of the primary calibrator and
the in-beam source.  Consistency between epochs was ensured by
constructing models of the primary and in-beam calibrators based on
all eight epochs of data, and using these models to iteratively
calibrate each individual epoch.

The astrometric positions and position uncertainties were fit for a
reference position, proper motion, and parallax using a linear least
squares algorithm. The random position uncertainties were determined
by fitting Gaussians to the pulsar images at each epoch and frequency.
Systematic errors were estimated by scaling the scatter in astrometric
positions at each epoch.  The scaling factor of 1.8 was determined
such that the final reduced $\chi^2$ for the fit was 1.0, with 59
degrees of freedom.  The fit results are summarized in \Tref{fit}, and
the proper motion and parallax signatures are plotted in \Fref{pmpi}.
We find a proper motion $\pmra = -73.606 \pm 0.044$ \mpy, $\pmdec =
-62.622 \pm 0.088$ \mpy, and a parallax $\pi = 0.415 \pm 0.037$ mas.
We verify the robustness of the parallax estimate by omitting each
epoch in turn from the fit and find  consistent results (within $\sim
1\sigma$).

\section{Distance, Velocity and Birth Site}\label{Sec:results}


From the measured parallax and proper motion, we calculate the
one-dimensional probability distributions for the distance ($D$) and
transverse velocity ($\vperp$). We infer the most probable values and
the most compact 68\% confidence intervals as $D =
2.37^{+0.23}_{-0.20}$~kpc and $\vperp = 1083^{+103}_{-90}$~\kms, the
first measured {\em model-independent} NS velocity $> 1000$~\kms.

The pulsar is at a large height above the Galactic plane ($z = D \sin
b \approx 2.0$~kpc), where DM based distance estimates are sensitive
to the model Galactic electron density scale height.  Based on the
pulsar DM of 19.6~pc~cm$^{-3}$, the model distance estimate is
$1.9^{+0.3}_{-1.2}$~kpc from TC93, and $1.0 \pm 0.2$~kpc from
NE2001. While TC93 is fortuitously close to the parallax distance,
NE2001 produces a significant underestimate, implying a much lower
electron density along part of the line of sight. Such a reduction may
possibly be due to voids or ``chimneys'' \citep{ni89} associated with
the Cygnus superbubble.  The electron density must then decay rapidly
after $\sim 0.8$~kpc along the line of sight.

While B1508+55 is well above the birth scale height for the Galactic
pulsar population \citep{acc02}, its proper motion vector ($\mu_l,
\mu_b = -6.39, 96.06$~\mpy\ after correcting for differential Galactic
rotation) suggests that it was born in the Galactic plane.  Given the
pulsar period $P$ and period derivative $\dot{P}$, and assuming that
the pulsar was born with a rapid initial spin and has been braking due
to a constant torque solely from magnetic dipole
radiation\footnote{While an estimate for $\ddot{P}$ exists for
this pulsar \citep{hlk+04}, it is affected by timing noise and does
not give a physically meaningful value for the braking index.} (i.e.,
$\dot{P} \propto P^{2-n}$, braking index $n = 3$), the spindown age
$\tau = P/2\dot{P} = 2.34$~Myr.

Tracing the pulsar back in the Galactic potential, we find that it
could have been born 2.34~Myr ago at $|z|<0.2$~kpc for a range of
modest (unknown) radial velocities between 0 and 300~\kms, and sample
orbits are shown in \Fref{orbit}.  The pulsar appears to have
originated in or around the Cyg OB associations, although we cannot
identify the exact birth cluster due to the lack of proper motions and
accurate distance estimates for the clusters.  Spindown age is known
to be an imperfect estimate of the pulsar chronological age
\citep[e.g.,][]{nv81,cc98}.  If the true age of the pulsar is
significantly larger than the spindown age, the pulsar birth location
would be several kpc beyond the Cyg OB associations in the Galactic
plane.  However, reasonable values of the pulsar radial velocity and a
range of pulsar ages (or equivalently, a range of braking indices or
birth periods) produce a birth location in the Galactic plane toward
the Cyg OB associations.


\begin{figure}[thf]
\plotone{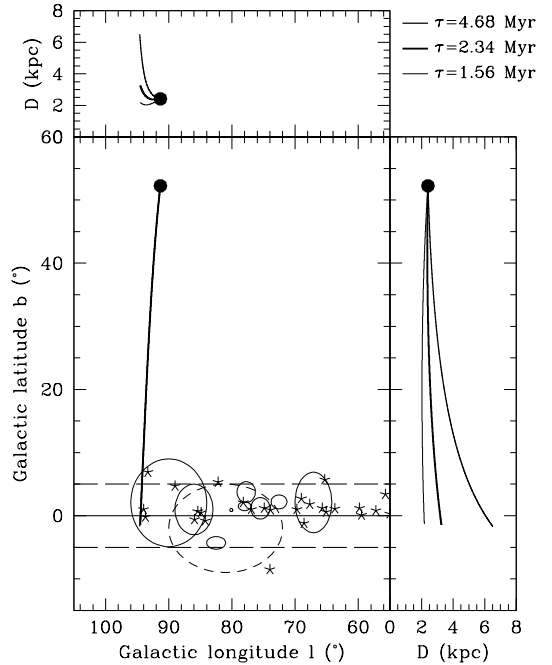}
\caption{Possible orbits for B1508+55, traced back in the Galactic
  potential.  The solid dot denotes the current pulsar position and
  the thick solid line the path it has followed for an age $\tau =
  2.34$~Myr (the spindown age, with braking index $n=3$) and a radial
  velocity $v_r = 200$~\kms. Other possible orbits are shown (thinner
  lines) for $\tau = 4.69$~Myr (i.e., $n=2$ for a small initial birth
  period) with corresponding $v_r = -300$~\kms, and for $\tau =
  1.56$~Myr ($n=4$) with $v_r = 700$~\kms.  Also indicated are the
  Cygnus superbubble (2~kpc away, dashed ellipse) and the Cyg OB
  associations (solid ellipses) with positions and extents as
  tabulated by \citet{ufr+01}.
  The starred symbols are the Galactic SN remnants identified in this
  region \citep{g04}. The solid horizontal line is the Galactic plane
  and the horizontal dashed lines indicate the pulsar birth scale height
  from \citet{acc02}
at the distance of the  Cygnus superbubble. 
}
\label{Fig:orbit}
\end{figure}

We thus establish a self-consistent narrative for B1508+55 that ties
together its current location, distance, and velocity vector with its
spindown age and birth site.  Born in the Cygnus region, B1508+55 is
moving rapidly away from the Galactic plane and on course to
irrevocably escape the Galaxy.  For $v_r = 200$~\kms, the implied
three-dimensional NS birth velocity $\vns \approx 1120$~\kms.

\section{Discussion: Implications of the High Velocity}\label{Sec:discuss}

The high birth velocity of B1508+55 constrains evolutionary scenarios
for the pulsar.  It could have been born in an isolated system, in
which case a large natal kick is required.  Alternatively, it could
have formed in a binary system that was disrupted by a first or
second SN.  If so, its companion could have been a longer lived (and
thus less massive) star, or a compact object.  However, in order to
attain the highest possible velocity solely from binary disruption,
B1508+55 should have formed in a first SN and remained bound to a
massive He star companion in a tight orbit that was subsequently
disrupted by the second SN \citep{it96}.  The net velocity of the
pulsar then derives from either one or two mass-loss events.  Given
the steeply declining initial mass function for stellar progenitors,
He cores are expected to have masses $\lesssim 16 \Msun$.  The highest
velocities are obtained for the smallest possible binary radius, which
corresponds to the He core filling its Roche lobe (i.e., a
semi-detached binary).  Such binary systems would be rapidly
circularized due to the interaction between the NS and the envelope of
the massive companion.  (However, B1508+55 has a canonical field
strength, and thus cannot have accreted much material.)  For a
circular orbit, $ \vns = \sqrt{{G \Mc}/{a (1+\Mns/\Mc)}}$,
%
%
where $\Mc$ is the companion mass, $a$ is the orbital radius, and $\Mns$
is the NS mass (assumed to be $1.4\,\Msun$).  When such a system is
disrupted by a symmetric SN explosion of the massive companion, the
fraction of the orbital velocity imparted to the NS depends on the
mass ratio $\Mns/\Mc$ \citep[e.g.,][]{boersma61}.  Using estimates for
the companion star radius \citep{p71} and the size of the Roche lobe
\citep{v93} for companion masses of 4, 8, and 16~$\Msun$, we find
\citep[following][]{it96} $\vns \approx$ 40, 560, and 1000~\kms, with
a possible small contribution from the prior systemic velocity
($\lesssim 100$~\kms).  Thus the disruption of a close binary is very
unlikely to solely account for the birth velocity of B1508+55, and a
natal (or post-natal) kick is indicated.

Natal kicks from asymmetric SN explosions are believed to be
ubiquitous \citep[e.g.,][]{fk97,pv99,lcc01}, but the driving mechanism
remains unclear.  Global asymmetric perturbations during SN core
collapse may produce hydrodynamically driven kicks \citep{bh96}. In
recent two-dimensional simulations \citep{spj+04}, velocities
comparable to that measured for B1508+55 have been obtained.  However,
the first full three-dimensional simulations of SN core collapse
\citep{f04} suggest that as the proto-NS moves through the convective
region, asymmetric downflows and neutrino emission damp out the
initial kick velocities, producing final kicks $\lesssim 200$~\kms.
Such kicks are insufficient to account for the observed velocity of
B1508+55, but the issue is not yet settled.

If convective instabilities in core collapse simulations fail to
reproduce large kick velocities, this may suggest the existence of
neutrino driven kicks, and require the presence of extreme magnetic
fields ($B > 10^{16}$~G) during core collapse \citep{jr99,al99}.  Such
a high $B$ may also allow a post-natal rocket effect to accelerate the
pulsar \citep{lcc01}.  However, a kick driven by the NS $B$-field
should produce alignment between the spin axis and the velocity vector
due to rotation averaging.  The spin axis may be inferred from the
polarization sweep of the radio pulse for some pulsars
\citep[][however, this is model dependent]{rc69}.  For B1508+55, the
pulse centroid polarization angle (after correcting for Faraday
rotation along the line of sight) is $161\arcdeg \pm 5\arcdeg$
\citep{drr99}, which differs from the proper motion angle by $\sim
71\arcdeg$ (or $\sim 19\arcdeg$ for emission in the orthogonal mode).
Assuming it survives future scrutiny, the measurement does not support
alignment of the spin axis and $\vns$.

We conclude that a kick is required to provide B1508+55 with its high
birth velocity, but that a single model is not uniquely preferred
based on our current understanding.  It is possible that binary
disruption and a natal kick act in concert for this object.  Such a
composite scenario would produce a misalignment between the spin axis
and velocity vector, as is (possibly) observed, with the relative
weights of the contributions depending on whether the emission is in
regular or orthogonal modes.  However, such a scenario is not well
supported unless the polarization angle is verified and a runaway
binary companion is identified in the future \citep[e.g.][]{vcc04}.
Finally, we note that in their seminal paper, \citet{dt92} argued that
magnetars had high space velocities $\sim 1000$~\kms\ due to kicks
that required intense magnetic fields ($B \gtrsim 10^{15}$~G), and
that such strong kicks would not be accessible to ordinary pulsars
(like B1508+55, with $B = 2 \times 10^{12}$~G).  If this argument is
to hold, then B1508+55 had a large birth magnetic field that rapidly
decayed.  It is more likely that $B$-driven kicks are not {\em
required} for high velocities, and that large natal kicks are not a
defining difference between pulsars and magnetars.  Rather, the
dividing line between them may lie elsewhere in their birth and
evolutionary history \citep[e.g., in their progenitor
mass;][]{gmo+04}.

Ongoing deep pulsar searches, and especially searches at high Galactic
latitudes, coupled with long-term VLBA astrometry programs, are likely
to discover further high-velocity pulsars and continue to elucidate
the velocity distribution and birth circumstances of neutron stars.

\acknowledgements
We acknowledge the Very Long Baseline Array operations team for their
efforts in scheduling and supporting a large VLBA astrometry
program. S. C. gratefully acknowledges support from the National Radio
Astronomy Observatory (NRAO) through a Jansky Fellowship.  NRAO is a
facility of the National Science Foundation operated under cooperative
agreement by Associated Universities, Inc.  Basic research in radio
astronomy at the Naval Research Laboratory is supported by the Office
of Naval Research.  This work was supported in part by NSF grants AST
98-19931 and AST 02-06036 at Cornell, and AST 00-98343 at the University
of California.


\begin{thebibliography}{38}
\expandafter\ifx\csname natexlab\endcsname\relax\def\natexlab#1{#1}\fi

\bibitem[{{Arras} \& {Lai}(1999)}]{al99}
{Arras}, P., \& {Lai}, D. 1999, \apj, 519, 745

\bibitem[{{Arzoumanian} {et~al.}(2002){Arzoumanian}, {Chernoff}, \&
  {Cordes}}]{acc02}
{Arzoumanian}, Z., {Chernoff}, D.~F., \& {Cordes}, J.~M. 2002, \apj, 568, 289

\bibitem[{{Blaauw}(1961)}]{b61}
{Blaauw}, A. 1961, \bain, 15, 265

\bibitem[{{Boersma}(1961)}]{boersma61}
{Boersma}, J. 1961, \bain, 15, 291

\bibitem[{{Brisken} {et~al.}(2002){Brisken}, {Benson}, {Goss}, \&
  {Thorsett}}]{bbgt02}
{Brisken}, W.~F., {Benson}, J.~M., {Goss}, W.~M., \& {Thorsett}, S.~E. 2002,
  \apj, 571, 906

\bibitem[{{Burrows} \& {Hayes}(1996)}]{bh96}
{Burrows}, A., \& {Hayes}, J. 1996, Physical Review Letters, 76, 352

\bibitem[{{Chatterjee} {et~al.}(2001){Chatterjee}, {Cordes}, {Lazio}, {Goss},
  {Fomalont}, \& {Benson}}]{ccl+01}
{Chatterjee}, S., {Cordes}, J.~M., {Lazio}, T.~J.~W., {Goss}, W.~M.,
  {Fomalont}, E.~B., \& {Benson}, J.~M. 2001, \apj, 550, 287

\bibitem[{{Cordes} \& {Chernoff}(1998)}]{cc98}
{Cordes}, J.~M., \& {Chernoff}, D.~F. 1998, \apj, 505, 315

\bibitem[{{Cordes} \& {Lazio}(2005)}]{cl02}
{Cordes}, J.~M., \& {Lazio}, T.~J.~W. 2005, ApJ, submitted; astro-ph/0207156

\bibitem[{{Cordes} {et~al.}(1993){Cordes}, {Romani}, \& {Lundgren}}]{crl93}
{Cordes}, J.~M., {Romani}, R.~W., \& {Lundgren}, S.~C. 1993, \nat, 362, 133

\bibitem[{{Deshpande} {et~al.}(1999){Deshpande}, {Ramachandran}, \&
  {Radhakrishnan}}]{drr99}
{Deshpande}, A.~A., {Ramachandran}, R., \& {Radhakrishnan}, V. 1999, \aap, 351,
  195

\bibitem[{{Dewey} \& {Cordes}(1987)}]{dc87}
{Dewey}, R.~J., \& {Cordes}, J.~M. 1987, \apj, 321, 780

\bibitem[{{Duncan} \& {Thompson}(1992)}]{dt92}
{Duncan}, R.~C., \& {Thompson}, C. 1992, \apjl, 392, L9

\bibitem[{{Fomalont} {et~al.}(1999){Fomalont}, {Goss}, {Beasley}, \&
  {Chatterjee}}]{fgbc99}
{Fomalont}, E.~B., {Goss}, W.~M., {Beasley}, A.~J., \& {Chatterjee}, S. 1999,
  \aj, 117, 3025

\bibitem[{{Fryer} \& {Kalogera}(1997)}]{fk97}
{Fryer}, C., \& {Kalogera}, V. 1997, \apj, 489, 244

\bibitem[{{Fryer}(2004)}]{f04}
{Fryer}, C.~L. 2004, \apjl, 601, L175

\bibitem[{{Gaensler} {et~al.}(2005){Gaensler}, {McClure-Griffiths}, {Oey},
  {Haverkorn}, {Dickey}, \& {Green}}]{gmo+04}
{Gaensler}, B.~M., {McClure-Griffiths}, N.~M., {Oey}, M.~S., {Haverkorn}, M.,
  {Dickey}, J.~M., \& {Green}, A.~J. 2005, \apjl, 620, L95

\bibitem[{{Gott} {et~al.}(1970){Gott}, {Gunn}, \& {Ostriker}}]{ggo70}
{Gott}, J.~R.~I., {Gunn}, J.~E., \& {Ostriker}, J.~P. 1970, \apjl, 160, L91

\bibitem[{{Green}(2004)}]{g04}
{Green}, D.~A. 2004, Bulletin of the Astronomical Society of India, 32, 335

\bibitem[{{Harrison} \& {Tademaru}(1975)}]{ht75}
{Harrison}, E.~R., \& {Tademaru}, E. 1975, \apj, 201, 447

\bibitem[{{Hobbs} {et~al.}(2005){Hobbs}, {Lorimer}, {Lyne}, \&
  {Kramer}}]{hllk05}
{Hobbs}, G., {Lorimer}, D.~R., {Lyne}, A.~G., \& {Kramer}, M. 2005, \mnras,
  360, 974

\bibitem[{{Hobbs} {et~al.}(2004){Hobbs}, {Lyne}, {Kramer}, {Martin}, \&
  {Jordan}}]{hlk+04}
{Hobbs}, G., {Lyne}, A.~G., {Kramer}, M., {Martin}, C.~E., \& {Jordan}, C.
  2004, \mnras, 353, 1311

\bibitem[{{Iben} \& {Tutukov}(1996)}]{it96}
{Iben}, I.~J., \& {Tutukov}, A.~V. 1996, \apj, 456, 738

\bibitem[{{Janka} \& {Mueller}(1996)}]{jm96}
{Janka}, H.-T., \& {Mueller}, E. 1996, \aap, 306, 167

\bibitem[{{Janka} \& {Raffelt}(1999)}]{jr99}
{Janka}, H.-T., \& {Raffelt}, G.~G. 1999, \prd, 59, 023005

\bibitem[{{Lai} {et~al.}(2001){Lai}, {Chernoff}, \& {Cordes}}]{lcc01}
{Lai}, D., {Chernoff}, D.~F., \& {Cordes}, J.~M. 2001, \apj, 549, 1111

\bibitem[{{Manchester} {et~al.}(2001){Manchester}, {Lyne}, {Camilo}, {Bell},
  {Kaspi}, {D'Amico}, {McKay}, {Crawford}, {Stairs}, {Possenti}, {Kramer}, \&
  {Sheppard}}]{pmb1}
{Manchester}, R.~N.~et~al.\ 2001, \mnras, 328, 17

\bibitem[{{Narayan} \& {Vivekanand}(1981)}]{nv81}
{Narayan}, R., \& {Vivekanand}, M. 1981, \nat, 290, 571

\bibitem[{{Norman} \& {Ikeuchi}(1989)}]{ni89}
{Norman}, C.~A., \& {Ikeuchi}, S. 1989, \apj, 345, 372

\bibitem[{{Paczy{\' n}ski}(1971)}]{p71}
{Paczy{\' n}ski}, B. 1971, Acta Astronomica, 21, 1

\bibitem[{{Portegies Zwart} \& {van den Heuvel}(1999)}]{pv99}
{Portegies Zwart}, S.~F., \& {van den Heuvel}, E.~P.~J. 1999, New Astronomy, 4,
  355

\bibitem[{{Radhakrishnan} \& {Cooke}(1969)}]{rc69}
{Radhakrishnan}, V., \& {Cooke}, D.~J. 1969, \aplett, 3, 225

\bibitem[{{Scheck} {et~al.}(2004){Scheck}, {Plewa}, {Janka}, {Kifonidis}, \&
  {M{\" u}ller}}]{spj+04}
{Scheck}, L., {Plewa}, T., {Janka}, H.-T., {Kifonidis}, K., \& {M{\" u}ller},
  E. 2004, Physical Review Letters, 92, 011103

\bibitem[{{Shklovskii}(1970)}]{s70}
{Shklovskii}, I.~S. 1969, \azh, 46, 715

\bibitem[{{Taylor} \& {Cordes}(1993)}]{tc93}
{Taylor}, J.~H., \& {Cordes}, J.~M. 1993, \apj, 411, 674

\bibitem[{{Uyan{\i}ker} {et~al.}(2001){Uyan{\i}ker}, {F{\" u}rst}, {Reich},
  {Aschenbach}, \& {Wielebinski}}]{ufr+01}
{Uyan{\i}ker}, B., {F{\" u}rst}, E., {Reich}, W., {Aschenbach}, B., \&
  {Wielebinski}, R. 2001, \aap, 371, 675

\bibitem[{{Verbunt}(1993)}]{v93}
{Verbunt}, F. 1993, \araa, 31, 93

\bibitem[{{Vlemmings} {et~al.}(2004){Vlemmings}, {Cordes}, \&
  {Chatterjee}}]{vcc04}
{Vlemmings}, W.~H.~T., {Cordes}, J.~M., \& {Chatterjee}, S. 2004, \apj, 610,
  402

\end{thebibliography}

\end{document}